\documentclass[preprint,aps,floats,showpacs,amssymb,tightenlines]{revtex4}
\usepackage{epsfig}

\newcommand{\ee}{\end{equation}}
\newcommand{\be}{\begin{equation}}

\newcommand{\nn}{\nonumber}

\renewcommand{\thefootnote}{\fnsymbol{footnote}}

\renewcommand{\thefootnote}{\arabic{footnote}}

\begin{document}
\title{Twisted semilocal strings in the MSSM}
\renewcommand{\thefootnote}{\fnsymbol{footnote}}
\vskip 1.5 cm
\author{Y. Brihaye\footnote{brihaye@umh.ac.be} \ {\small and} \   L. Honorez}
\affiliation{Physique Th\'eorique et Math\'ematique, \\ 
Universit\'e de Mons-Hainaut, Place du Parc, \\
B-7000, Mons, Belgique.}
\date{\today}
\begin{abstract}
The standard electroweak model is extended by means of a second
Brout-Englert-Higgs-doublet. The symmetry breaking potential
is chosen in such a way that (i) the Lagrangian possesses a custodial
symmetry, (ii) a stationary, axially symmetric
ansatz of the bosonic fields consistently reduces the Euler-Lagrange
equations to a set of differential equations.
The potential involves, in particular,
a direct interaction between the two doublets.
Stationary, axially-symmetric solutions of the classical equations
are constructed. Some of them can be assimilated to embedded Nielsen-Olesen strings.
From these solutions there are bifurcations and new solutions appear which exhibit
the characteristics of the recently constructed twisted semilocal strings.
A special emphasis is set on ``doubly-twisted" solutions for which the two doublets present different 
time-dependent phase factors. They are regular and have
a finite energy which can be lower than the energy of the embedded twisted solution.
Electric-type  solutions, such that the fields oscillate asymptotically
far from the symmetry-axis, are also reported.   
\end{abstract}
\maketitle
\medskip \medskip
\newpage
\section{Introduction}
Since the discovery of topological string solutions in the so called ``Abelian-Higgs-model''
 by Nielsen-Olesen  \cite{no},
more than thirty years ago, there have been several investigations for applying these solutions
in high energy physics and cosmology. The existence of these solutions is closely connected to
the symmetry breaking of the local U(1)-symmetry by a Brout-Englert-Higgs mechanism. 
In the last years it has been shown that  string-like configurations also appear
as classical solutions in several other models within several pattern of symmetry breaking.  One
of the most famous generalisation of the string is the  ``semi-local'' string \cite{vachaspati}. 
It appears in models involving both local and global symmetry breaking, see \cite{achucarro} for a review.  
One of the most striking properties of these solutions is that they constitute stable vortices without being
supported by a standard topological argument related to the  homotopy group of the vacuum manifold \cite{hindmarsh0}.  
The semi-local string appears naturally in models involving N charged complex scalar fields coupled to electromagnetism
and presenting an SU(N) global symmetry suitably broken by a Brout-Englert-Higgs potential.
Along with the basic Nielsen-Olesen (NO) strings they possess a cylindrical symmetry, they do not
depend on the coordinate associated to the symmetry axis (usually chosen as the $oz$-axis) and have
a finite energy per unit length of the $z$ coordinate.

Very recently, the classical equations of this model (in the case N=2) were reinvestigated \cite{fv1,fv2}
and new families of solutions were shown to exist. These remarkable solutions
 are called ``twisted-semilocal strings''.  The  
crucial difference between the twisted semilocal and the conventional ones \cite{achucarro,vachaspati}
 resides in the fact
that the two scalar fields have different phase factors depending linearly on time as well as on the coordinate
corresponding to their axis of symmetry. This needs an axially symmetric ansatz but, e.g. does not invalidate 
the cylindrical symmetry of the energy density of these solutions.

In this paper, we consider the semi-local strings (twisted or not) in the framework of a model involving
two Brout-Englert-Higgs doublets. This is motivated by the Minimal Supersymmetric Standard Model (MSSM).
Along with \cite{fv1,fv2}, we consider the classical equations in the limit where the Weinberg angle $\theta_W$
is set to $\theta_W = \pi/2$.   
The most general potential used in the context of MSSM is very involved and depends roughly on ten parameters.
Here we limit our investigation to the subfamily of parameters allowing for a custodial symmetry. This restricts
the number of parameters to five.
We give numerical evidence that, in a specific but rather large region of the space of the parameters (involving
to a large extent the physical region determined by the experimental lower bounds for the Brout-Englert-Higgs bosons)
solutions exist for which the two doublets present independent twists in the phases associated with specific
components of them. We call this type of solutions  ``doubly-twisted'', in general they coexist with
the embedded simply-twisted solutions \cite{fv1,fv2} where one of the doublets does not
depend on $t$ and $z$. For some region of the parameters, doubly twisted solutions have the lowest energy.
The parameters labeling the twist can, in particular, be tuned in such a way
that the associated Noether currents have opposite signs. Another feature of 
doubly twisted solutions is that, for generic
values of  the potential, the components $A_0$ and $A_z$ of the Maxwell field turn out to
be not proportional to each other. This contrasts with the case of one doublet \cite{fv2}.

The description of the model is presented in Sect. II, where a special emphasis is set to the parameters
of the potential and their usual relations to the masses of the Higgs-particles. The ansatz for the fields
and the corresponding equations are given is Sect. III. In Sect. IV, we characterize the constraints of the
potential that allow embedded NO strings. This is useful since it leads to a family
of semi-analytical solutions (in a sense that they are  parametrized in terms of the profile of the NO solution only). 
Several features of the solutions constructed  numerically for generic values of the potential are 
discussed in Sect. V. Due to the possibilities allowed by the twist with respect to the two doublets,
the solutions depend on several independent parameters. We present a few plots which hopefully capture
a large part of qualititive properties of the twisted semi-local string available in the MSSM. Some
conclusions are given in Sect. VI. 
\section{The model}
The lagrangian that we consider in this paper reads
\be
\label{lag}
{\cal L} = -\frac{1}{4} F_{\mu\nu}^a F^{\mu\nu,a} + -\frac{1}{4} f_{\mu\nu} f^{\mu\nu}
+ (D_\mu \Phi_{(1)})^{\dagger} (D^\mu \Phi_{(1)})
+ (D_\mu \Phi_{(2)})^{\dagger} (D^\mu \Phi_{(2)})
- V(\Phi_{(1)}, \Phi_{(2)})
\ee
Where $\Phi_{(1)}, \Phi_{(2)}$ denote the two BEH-doublets
and the standard definitions are used  for the covariant derivative
and gauge field strengths~:
\be
\label{cova}
F_{\mu\nu}^a=\partial_\mu V_\nu^a-\partial_\nu V_\mu^a
            + g \epsilon^{abc} V_\mu^b V_\nu^c
\ee
\be
D_{\mu} \Phi = \Bigl(\partial_{\mu}
             -\frac{i}{2}g \tau^a V_{\mu}^a - \frac{i}{2}g' A_{\mu}
               \Bigr)\Phi
\ee
(the case $\theta_W=\pi/2$, i.e. $g=0$ will be used later in this paper).

The most general gauge invariant potential constructed with 
two BEH-doublets
is presented namely in \cite{sophie,hunter} and depends on nine coupling constants.
Here we consider  the family of potentials of the form
\be
\label{pot}
     V(\Phi_{(1)}, \Phi_{(2)})
    = \lambda_1 (\Phi_{(1)}^{\dagger} \Phi_{(1)} - {v_1^2 \over 2})^2
    + \lambda_2 (\Phi_{(2)}^{\dagger} \Phi_{(2)} - {v_2^2 \over 2})^2
    + \lambda_3  (\Phi_{(1)}^{\dagger} \Phi_{(1)} - {v_1^2 \over 2}) 
    (\Phi_{(2)}^{\dagger} \Phi_{(2)} - {v_2^2 \over 2})
\ee
depending on five parameters. 
The direct coupling between the two doublets is parametrized
by the constant $\lambda_3$
and one of  the main characteristic about this  potential resides 
in the fact that
it imposes a symmetry breaking mechanism to each of the BEH-doublets.
Several classical solutions in the models have been studied in the past.
Spherically symmetric sphaleron solutions were studied at length
in the case $\lambda_3 = 0$ in \cite{btt,kleihaus,hindmarsh} and a 
similar study for $\lambda_3 \neq 0$ was reported in \cite{brihaye}.

The lagrangian (\ref{lag}) is of course invariant under SU(2) gauge
transformations but it further possesses a larger global symmetry
under SU(2) $\times$ SU(2) $\times$ SU(2).
In fact, the part of the lagrangian (\ref{lag}) involving
the scalar fields  can be
written in terms of 2$\times$2 matrices defined by
\be
      M_{1,2} \equiv
       \left(\matrix {\phi_0^*& \phi_+\cr
                      -\phi_+^* & \phi_0\cr}\right)_{1,2}
                \ \ \ {\rm for} \ \ \
  \Phi_{1,2} =  \left(\matrix {\phi_+ \cr \phi_0\cr } \right)_{1,2}
\ee
When written  in terms of the matrices $M_1$ and $M_2$, the lagrangian
(\ref{lag}) becomes manifestly invariant under the transformation
\be
\label{custo}
   V_{\mu}' = A V_{\mu} A^{\dagger} \ \ , \ \
   M_1' = A M_1 B  \ \ , \ \
   M_2' = A M_2 C
\ee
 with $A,B,C \in$ SU(2); this is the custodial symmetry.
The double symmetry breaking mechanism imposed by the potential
(\ref{pot}) leads to a mass $M_W$ for two of the three gauge
vector bosons and, namely, to two neutral BEH-particles
with masses $M_h$, $M_H$.
In terms of the parameters of the Lagrangian, these masses are
given by  \cite{hunter}
\be
\label{masse}
     M_W = {g \over 2} \sqrt{v_1^2 + v_2^2} \ \ \ , \ \ \
M^2_{H,h} = \frac{1}{2}[A_1 + A_2 \pm \sqrt{(A_1-A_2)^2 + 4 B^2}]
\ee
with
\be
    A_1 = 2 v_1^2(\lambda_1 + \lambda_3) \ \ , \ \
    A_2 = 2 v_2^2(\lambda_2 + \lambda_3) \ \ , \ \
    B = 2 \lambda_3 v_1 v_2
\ee
For later convenience we also define
\be
\label{param}
    \tan \beta = {v_2 \over v_1} \ \ , \ \ 
    \rho_{H,h} = {M_{H,h} \over M_W}  \ \ , \ \
\epsilon_p = 4 {\lambda_p \over g^2} \ \ , \ \ p=1,2,3
\ee
Note that the mass ratios $\rho_{1,2}$ used in \cite{kleihaus}
are related to $\rho_{H,h}$ by 
$\rho_H = {\rm max}\{\rho_1,\rho_2\}$,
$\rho_h = {\rm min}\{\rho_1,\rho_2\}$.
For physical reasons,  we consider only $v_1 \geq 0$, $v_2 \geq 0$
so that $0 \leq \beta \leq \pi/2$. Interestingly, the parameter 
$\epsilon_3$ can be negative but cannot take arbitrary values. 
The following relations are useful to determine the physical region~:
\begin{eqnarray}
\epsilon_1 \cos^2 \beta + \epsilon_2 \sin^2 \beta 
&=& \frac{1}{2}(\rho_H^2 + \rho_h^2) -  \epsilon_3 \nonumber \\
\epsilon_1 \cos^2 \beta - \epsilon_2 \sin^2 \beta 
 &=& \frac{1}{2} \sqrt{(\rho_H^2 - \rho_h^2)^2 - 4 \epsilon_3^2 \sin^2(2 \beta)}
- \epsilon_3 \cos(2 \beta)
\end{eqnarray}
The physical domain is then determined by the conditions
\begin{equation}
\frac{\rho_h^2-\rho_H^2}{2 \sin(2\beta)} \leq \epsilon_3   \leq
\frac{\rho_H^2-\rho_h^2}{2 \sin(2 \beta)} \ \ \ \  , \ \ \ \ 
              0 \leq  \rho_h^2 \leq \rho_H^2
\end{equation}

 \section{Cylindrical symmetry}
 In this paper, we will assume that only the U(1)-subgroup of the gauge group is really
 gauged (that is to say that we assume $V_{\mu}^1 = V_{\mu}^2 =0$)  and look for solutions 
 presenting a cylindrical symmetry. For this purpose, we impose a
form for the fields inspired from \cite{fv1,fv2}. After some algebra, it turns out that
the form
\be
    \phi_{1(1)} = f_1 (\rho) \exp^{i n_1\phi}   \ \ , \ \    
    \phi_{2(1)} = f_2 (\rho) \exp^{i m_1\phi} \exp^{i(\omega_0 t + \omega_3 z)}
\ee
\be
    \phi_{1(2)} = g_1 (\rho) \exp^{i n_2\phi} \exp^{i(\sigma'_0 t + \sigma'_3 z)}  \ \ , \ \    
    \phi_{2(2)} = g_2 (\rho) \exp^{i m_2\phi} \exp^{i(\sigma_0 t + \sigma_3 z)}
\ee
\be
\label{ansatz}
A_0 = \omega_0 a_0(\rho) \ \ , \ \ A_{\rho} = 0 \ \ , \ \ A_{\phi} = n a(\rho) \ \ , \ \ A_3 = \omega_3 a_3(\rho) \ \ ,
\ee
 $n_1,n_2$ and $m_1,m_2$ being integers is consistent with the classical equations.As pointed out in \cite{fv1,fv2}, solutions exist for $0\leq{m_1}\leq{n_1}$ and $0\leq{m_2}\leq{n_2}$. Inserting this ansatz into the energy functional, it turns out that finite energy configurations are possible only for $n_1=n_2=n$ and $\sigma'_0=\sigma'_3=0$. 
In fact we parametrize  the U(1) field as close as possible to \cite{fv1,fv2} (although without losing generality)
 in order to crosscheck our solutions with the 
solutions constructed of \cite{fv1,fv2}  in the following two limits~: (i) the  second doublet 
$\Phi_{(2)}$ becomes trivial, (ii) the parameters $\sigma_{0,3}$ appearing in this doublet vanish.

The ansatz above leads to a reduced lagrangian density of the form~: 
 \begin{eqnarray}
 {\cal{L}} &=&  \rho \ \frac{1}{2}\left[\omega^2_0(a'_0)^2  
 - \omega^2_3 (a'_3)^2  -  \frac{1}{\rho^2}n^2(a')^2 \right] \nn \\
   &-&  \rho ( \cos\beta)^2   [ (f'_1)^2  +  (f'_2)^2 + (g'_1)^2  +  (g'_2)^2
   + ( - \omega^2_0 a^2_0 + \omega^2_3 a^2_3
   + \frac{n^2 (1-a)^2}{\rho^2})f^2_1  \nn \\
   &+& ( - \omega^2_0 ( 1- a_0)^2 +  \omega^2_3 ( 1- a_3)^2
  +\frac{(m_1 - na)^2}{\rho^2} )f^2_2
   + ( - \omega^2_0 a^2_0 + \omega^2_3 a^2_3
   + \frac{n^2 (1-a)^2}{\rho^2})g^2_1 \nn \\
   &+&  ( -(\sigma_0 - \omega_0a_0)^2  +  (\sigma_3 - \omega_3a_3)^2
  +  \frac{(m_2 - na)^2}{\rho^2} ) g^2_2 ] \nn \\
   &-&  \rho (\cos\beta)^4 \left[ \epsilon_1 ( |f|^2 - 1)^2
 +   \epsilon_2 ( |g|^2  - {\rm tg} ^2\beta )^2
  +  \epsilon_3( |f|^2 - 1) ( |g|^2  -  {\rm tg}^2\beta )\right]
\end{eqnarray}
where the prime means the derivative with respect to the axial variable $\rho$.
The quantity $Q$ defined according to
\be
\label{q}
      Q \equiv 2 \pi \int d\rho \rho a_3 (f_1^2 + g_1^2) = 2 \pi \int d\rho \rho (1-a_3) (f_2^2 + g_2^2)
\ee
is also of great interest. It
determines the vortex worldsheet currents and generalizes Eqs.(5.4)-(5.6) of \cite{fv2}.
They are associated with the conserved quantities related to the Noether current of the residual U(1) symmetry. 
In passing we note that the equality in (\ref{q}) provides a useful test of accuracy for our numerical solutions.


In order to write the corresponding Euler-Lagrange equations, we find it convenient
to define  $K =  (a_0 + a_3)/2$, $\Delta =  (a_0 - a_3)/2$. It is straightforward to establish 
the Euler-Lagrange 
equations~:
\begin{eqnarray}
 \rho (\frac{a'}{\rho})' &=& 2(\cos \beta)^2 \left\{ f^2_1 (a - 1) + f^2_2 (a - \frac{m_1}{n})  + g^2_1 (a - 1) + g^2_2 (a - \frac{m_2}{n}) \right\} \nn\\
\frac{1}{\rho} (\rho f'_1)' &=& f_1 \left\{  - \omega^2_0 a^2_0 + \omega^2_3 a^2_3 + \frac{n^2 (1-a)^2 }{\rho^2}\right\}\nn\\
&&\qquad \qquad
 + (\cos \beta)^2 \left\{   2 \epsilon_1 f_1 (|f|^2 - 1) + \epsilon_3 f_1 (|g|^2 - {\rm tg}^2\beta) \right\}\nn\\
\frac{1}{\rho} (\rho f'_2)' &=& f_2 \left\{  - \omega^2_0 (1 - a_0)^2 + \omega^2_3(1-  a_3)^2 + \frac{(m_1 - na)^2}{\rho^2}\right\}\nn\\
&&\qquad \qquad + (\cos \beta)^2 \{ 2 \epsilon_1 f_2 (|f|^2 - 1) + \epsilon_3 f_2 (|g|^2 - {\rm tg}^2\beta) \}\nn\\
\frac{1}{\rho} (\rho g'_1)' &=& g_1 \left\{  - \omega^2_0 a^2_0 + \omega^2_3 a^2_3 + \frac{n^2 (1-a)^2}{\rho^2}\right\} \nn\\
&&\qquad \qquad + (\cos \beta)^2\left\{ 2 \epsilon_2 g_1 (|g|^2 - {\rm tg}^2\beta) + \epsilon_3 g_1 (|f|^2 -1) \right\}\nn\\
\frac{1}{\rho} (\rho g'_2)' &=& g_2 \left\{  - (\sigma_0 - \omega_0 a_0)^2 + (\sigma_3 - \omega_3 a_3)^2 + \frac{(m_2 - na)^2}{\rho^2}\right\}\nn\\
&&\qquad \qquad + (\cos \beta)^2\left\{ 2 \epsilon_2 g_2 (|g|^2 - {\rm tg}^2\beta) + \epsilon_3 g_2 (|f|^2 - 1) \right\}\nn\\
\frac{1}{\rho} (\rho K')' &=& (\cos \beta)^2 \left\{ 2K |f|^2 - 2 f^2_2 +  2K |g|^2 - g^2_2 (\frac{\sigma_0}{\omega_0} + \frac{\sigma_3}{\omega_3}) \right\}\nn\\
\frac{1}{\rho} (\rho \Delta')' &=& (\cos \beta)^2 \left\{ 2\Delta |f|^2 +  2\Delta |g|^2 - g^2_2 (\frac{ \sigma_0}{\omega_0} - \frac{\sigma_3}{\omega_3}) \right\}
\end{eqnarray}
where $|f|^2 \equiv f^2_1 + f^2_2$  and  $|g|^2 \equiv g^2_1 + g^2_2$.
The interest for using $\Delta$ appears on the last equation; defining 
$R_0 \equiv (\sigma_0/\omega_0)$ , $R_3 \equiv (\sigma_3/\omega_3)$ and 
$\delta \equiv R_0-R_3$, we see that $\delta = 0$  
(corresponding to parallel twists in the two doublets)
 implies $\Delta=0$ by the positivity argument. If however solutions exist such that 
$ \delta \neq 0$, they will have
$\Delta \neq 0$; this is a new feature with respect to the case with one doublet \cite{fv1,fv2}.

In order to obtain regular and finite energy solutions, the above system has to be solved with the following
boundary conditions respectively at $\rho=0$ and $\rho=\infty$~:
\be
\label{boundary1}
  a_0'(0) = a_3'(0) = a(0) = f_1(0)= f_2'(0) = g_1(0) = g_2'(0) = 0 
\ee
\be
\label{boundary2}
  a_0(\infty) = a_3(\infty) = f_2(\infty) = g_2(\infty) = 0 \ \ ,  
  a(\infty)=1 , \ f_1(\infty)=1  , \  g_1(\infty) = \tan(\beta) 
\ee
Before discussing the solutions for generic values of the potential's parameters, it is useful to 
study a few specific limits of these equations.
\section{Particular solutions}
\subsection{Nielsen-Olesen strings}
Setting $\epsilon_2=\epsilon_3=\beta =0$ in the potential and $a_3=a_0=f_2=g_2=g_1=0$, the
system of equations 
above naturally reduces to the two equations~:
\be
   a'' - \frac{1}{\rho} a' = 2f^2(a-1) \ \ \ , \ \ \ f'' + \frac{1}{\rho}f'=f\frac{n^2}{\rho^2}(1-a)^2 + 2 \epsilon f(f^2-1)
\ \  , \  \epsilon = \epsilon_1
\ee
whose solution is the NO string. For later convenience, we denote this particular solution 
by $\bar a_{\epsilon}$, $\bar f_{\epsilon}$.
\subsection{Embedded Nielsen-Olesen string}
The system of seven equations above admits embedded NO solutions of the form
\be
    a = \bar a_{\epsilon} \ \ , \ \ f_1 = \bar f_{\epsilon}  \ \ , \ \ g_1 = \tan(\beta) \bar f_{\epsilon} \ \ , \ \ a_0 = a_3 = f_2 = g_2 = 0
\ee
provided the following relations between the parameters of the potential hold
\be
\label{constaint}
      2\epsilon = 2 \epsilon_1 \cos^2 \beta + \epsilon_3 \sin^2 \beta \ \ \ , \ \ \
      2\epsilon = 2 \epsilon_2 \sin^2 \beta + \epsilon_3 \cos^2 \beta \ \ \ .
\ee
Accordingly, embedded semilocal strings will exist if $\epsilon > 1/2$. 
For coupling constants away from these constraints, we expect the solutions to be deformed progressively from
the embedded NO-solutions.
 The linearized equations \cite{hindmarsh0} 
associated  with the fields $f_2$ and $g_2$ corresponding to an embedded NO-solution
lead to a Schroedinger equation
\be
     -\frac{1}{\rho} (\rho f_2')' + [\frac{1}{\rho^2}(n\bar a - m)^2 - 2\epsilon (1-\bar f^2)] f_2 = - \omega^2 f_2
     \ \ , \ \ m = m_1 = m_2
\ee
and an identical equation for $g_2$ with the spectral parameter $\omega^2$ replaced by $\sigma^2$.
Using the reasoning of \cite{fv1,fv2}, we conclude that superconducting solutions with
 both fields $\Phi_{(1)}$ and $\Phi_{(2)}$ twisted will exist as soon as the potential parameters 
 are such that $2 \epsilon > 1$. These will exist for
 \be
        \omega^2 \in ]0 , \omega_b^2] \ \ , \ \  \sigma^2 \in ]0 , \omega_b^2]
 \ee
where $\omega_b^2$ are computed in \cite{fv2} (see e.g. Tables I,II and Fig. 3 in \cite{fv2}). 
As a consequence, we expect solutions to exist with arbitrary relative phases associated with
the two Higgs doublets. This is indeed confirmed by our numerical analysis. See e.g. Fig. 4 where 
the energy, the charge $Q$ and the parameters $\Delta(0), f_2(0)$ are reported
as functions of the ratio $\sigma_3/\omega_3$, for $\omega_3=0.017$.

It is a natural question to ask whether the bifurcation point corresponding to $2 \epsilon=1$ in (\ref{constaint})
belongs to the physical region of the parameter of the MSSM.
In terms of the mass ratios $\rho_H,\rho_h$, it turns out that embedded NO-solutions exist if
\be
       \rho_H^2 + \rho_h^2 = 2 + \epsilon_3 \ \ , \ \ |\epsilon_3| / (4+ 5 \cos(2\beta)^2)^{1/2} = |\rho_H^2 - \rho_h^2|
\ee
If we fix $\rho_h$, the conditions above define a point
\be
  \rho_H = \sqrt{2 + \epsilon_3 - \rho_h^2}   \ \ , \ \   \epsilon_3 = \frac{2\sigma(1-\rho_h^2)}
  {(4+ 5 \cos^2(2\beta))^{1/2}-\sigma} \ \ , \ \ \sigma \equiv {\rm sign}(\epsilon_3)
\ee
in the $\rho_H, \epsilon_3$ plane. 
After some algebra, one can check that, if $\rho_h < 1$, this point is inside the physical
region for generic values of $\beta$ and {\it on} the curve determining the physical region for $\beta = \pi/4$.
Since the experimental results show that the Higgs mass is larger than the W-boson mass, no NO-embeded solution is available for realistic values of the Higgs mass parameter. 
\subsection{Embedded and mixed semi-local strings}
The twisted semi-local strings of \cite{fv1,fv2}
are characterized by the winding number $n$ and form $n$ families of solutions
labelled by $m$ with $0 \leq m \leq n-1$. Around $\rho=0$, the two components of the Higgs field behave 
according to $f_1 \sim f_1^{(n)} \rho^n$ and $f_2 \sim f_2^{(m)} \rho^m$.
When the condition (\ref{constaint}) holds, these solutions 
can also be embedded in the system above provided $g_1 = f_1\tan{\beta}$ and $g_2 = f_2 \tan{\beta}$
and $\delta = 0$, implying $\Delta(\rho)=0$ as
already mentionned. 
These embeddings holds for any $n$ and $m_1 = m_2$. 
However, it is likely that solutions can be constructed with $m_1 \neq m_2$ and $\delta=0$,
we will see that this is indeed the case and call these solutions `mixed'.

\subsection{Simply and doubly twisted solutions}
For generic values of the potential, the system of seven equations above admits solutions
where all the functions are non trivial and the parameters $R_0,R_3$ are independant. We will
denote these solutions doubly twisted ones. However, as can be seen from the equations
and the boundary conditions, solutions possibly exist which have $g_2=0$ and $f_2 \neq 0$ 
(or $f_2=0$ and $g_2 \neq 0$). In the following chapter, we will denote such solutions
simply twisted.  

\section{Numerical results}
The system above cannot be solved analytically, we relied on numerical method to construct solutions;
the numerical routine used  is based on the collocation method \cite{colsys}. 
Because both twisted and untwisted solutions obey the same set of boundary conditions, for instance 
(\ref{boundary1}), (\ref{boundary2}), 
some  trick has to be implemented in order to enforce the numerical routine to produce the twisted strings.
For this purpose we supplement the system
by three trivial equations, namely $d \omega_3/ d\rho = d\sigma_0/d \rho = d \sigma_3/d\rho = 0$ and we 
take advantage of these supplementary equations to impose by hand values for $f_2(0),g_2(0)$ and $a_3(0)$
 as supplementary boundary conditions. If we start from an appropriate initial configuration,
  the values of $\omega_3,\sigma_{0}, \sigma_3$ for which these supplementary
 boundary conditions are fullfilled can be reconstructed numerically.
  Once such a solution is available, the trick allows to 
 ``navigate'' very efficiently into the manifold of simply or doubly twisted solutions.

 \subsection{One doublet}
In order to test the efficiency of our technique, we first reconsidered the one-Higgs equations
in the case $\epsilon_1 = 2$, $\epsilon_{2,3}=0$ (corresponding to $\beta = 4$ in the notation of \cite{fv2}).
The second doublet then vanishes identically and the equations reduce to the ones of \cite{fv1,fv2}).
\begin{figure}[!htb]
\centering
\leavevmode\epsfxsize=10.0cm
\epsfbox{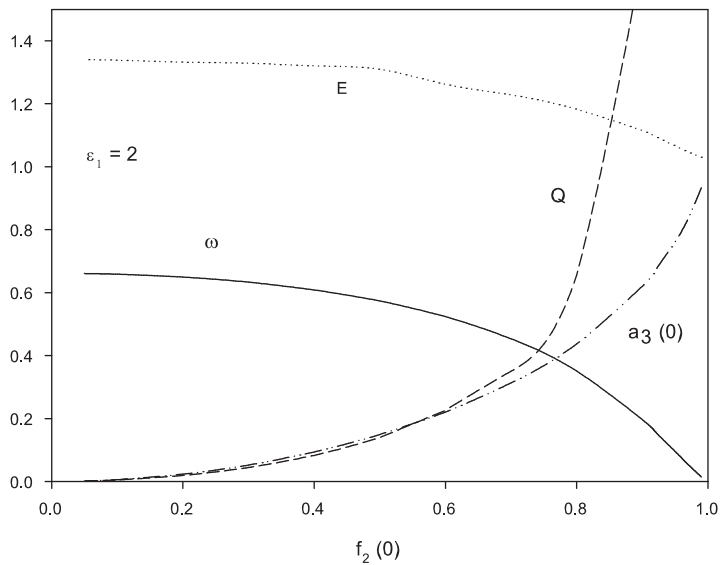}\\
\caption{\label{fig0}
 The energy, the charge $Q$  and parameters $a_3(0), \omega$ are plotted as 
 functions of the value $f_2(0)$  for the solution with one doublet and $\epsilon_1=2$.
}
\end{figure} 
The numerical values
that we obtain are in full agreement with those of \cite{fv2}.   
Several parameters characterizing this family of solutions are represented on Fig.\ref{fig0}. 
\subsection{Embedded solutions}
Assuming $R_3=R_0$, we manage to embed the solutions of \cite{fv2} in the MSSM model discussed above
i.e. solutions with $f_1=g_1, f_2=g_2$.
On Fig. 2, we present several parameters characterizing these embedded solutions for $\epsilon_1 = \epsilon_2=2$
$\epsilon_3 = 0$ and $\beta = \pi/4$ and for the solutions corresponding to
 $n=1 , \ m_1=m_2=0$; $n=2 , \ m_1=m_2=0$ and $n=2 , \ m_1=m_2=1$.
On these plots, the different parameters are represented as functions of $\omega$. The plots confirm
that the embedded solutions (like the original ones) exist only for a finite interval of $\omega$
and that the solution bifurcates into an embedded NO solution 
(with $f_2=g_2=0=a_3$) in the limit $\omega \to \omega_b$. In the limit $\omega \to 0$, a so called
Skyrme type solution is approached but we did not study this limit in details. We just note
that the charge $Q$ becomes infinite in this limit, as a consequence of the fact that the function $a_3$
reaches its asymptotic value $a_3(\infty)=0$ more and more slowly (see \cite{fv2} for more details).

\begin{figure}[!htb]
\centering
\leavevmode\epsfxsize=15.0cm
\epsfbox{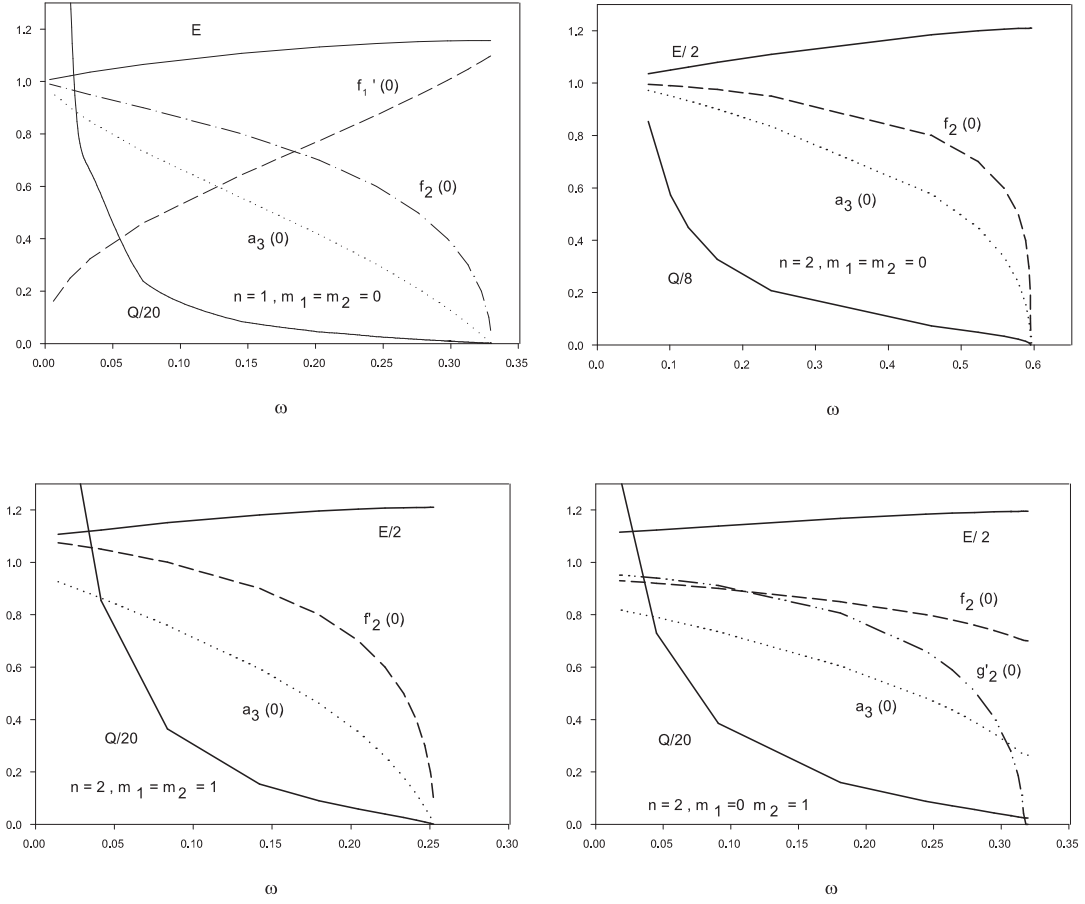}\\
\caption{\label{fig2}
 The energy, the charge $Q$  and parameters $a_3(0),f_1(0),f_2'(0)$, etc...  are plotted as 
 functions of $\omega$  for the embedded solutions corresponding to a few values of $n,m_1,m_2$ .
}
\end{figure} 
\subsection{Mixed solutions}
\begin{figure}[!htb]
\centering
\leavevmode\epsfxsize=10.0cm
\epsfbox{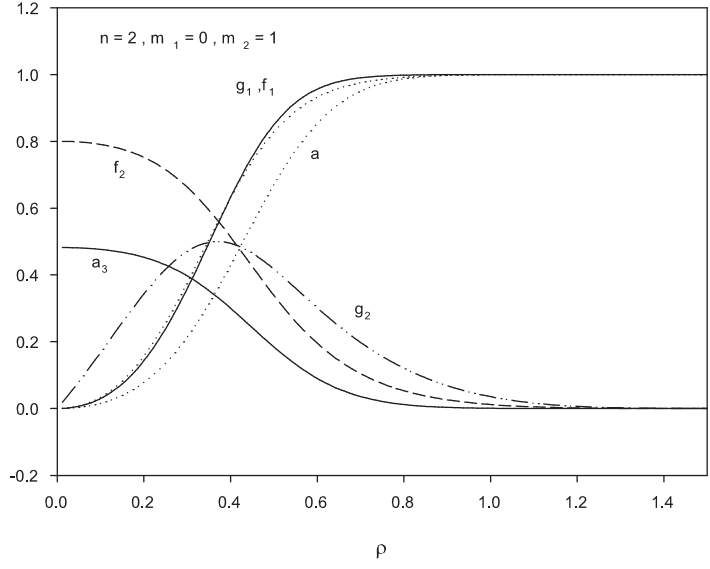}\\
\caption{\label{fig3}
 The profile of a mixed solution corresponding to $n=2,m_1=0,m_2=1$.
}
\end{figure}

The profile of a mixed a solution is presented on Fig. \ref{fig3} corresponding to $n=2,m_1=0, m_2=1$,
 $\delta=0$  and  $f_2(0)=0.8$; a branch of solutions of this type can be constructed by varying $f_2(0)$
A few parameters characterizing this branch with respect to $\omega$ are represented in Fig. 2. 
For the parameters choosen, it exists for $\omega < 0.32$. In the limit $\omega \to 0.32$, we have
$g_2'(0) \to 0$ and the function $g_2$ tends uniformly to zero while $a_3,f_2$ remain non zero, this feature is
specific to mixed solutions.  
\subsection{Doubly twisted solutions}
\begin{figure}[!htb]
\centering
\leavevmode\epsfxsize=10.0cm
\epsfbox{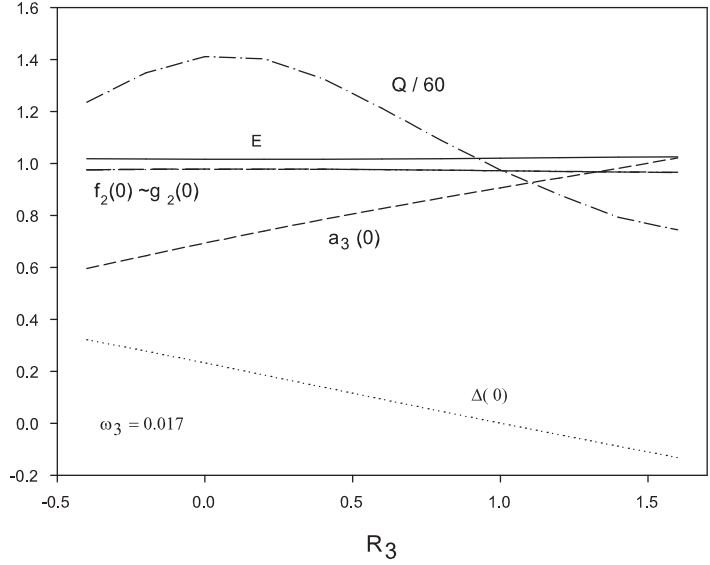}\\
\caption{\label{fig1}
 The energy, the charge $Q$  and parameters $\Delta(0),a_3(0), f_2(0)$ are plotted as functions of the ratio 
 $R_3$ for $R_0=1$ and $\omega_3 = 0.017$.
}
\end{figure}

We first investigate the effects of a doubly twisted solution of the NO-string type corresponding to
 $\epsilon_1=\epsilon_2=2$, $\epsilon_3=0$, $\beta = \pi/4$ and assuming $\omega_3 = 0.017$ , $R_0=1$. The effect
 of the parameter $R_3$ on the solutions is reported on Fig. \ref{fig1}. Remark 
 that for $R_3=1$, we have  $ \Delta(0)=0$; in fact we have  $\Delta(\rho)=0$ as a consequence 
 of the fact that the positivity argument applies  in the
 equation for the function $\Delta$ if $R_0=R_3$ .

\begin{figure}[!htb]
\centering
\leavevmode\epsfxsize=10.0cm
\epsfbox{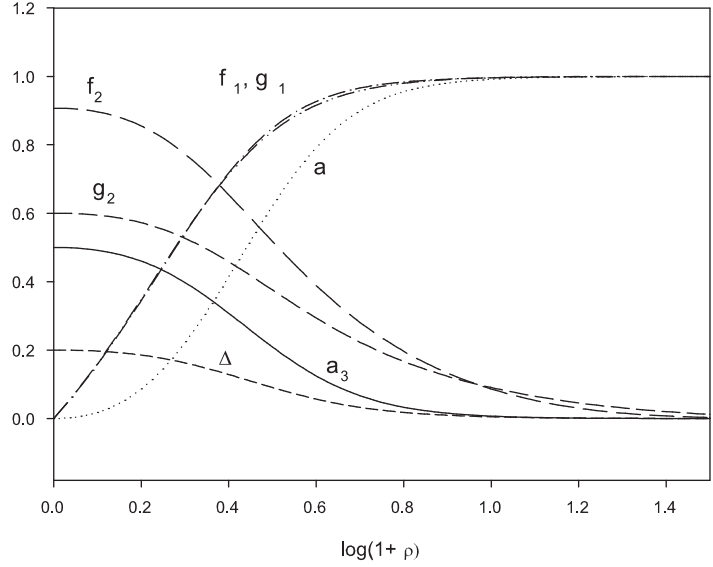}\\
\caption{
\label{profile} 
The profile of a generic solution corresponding to $\rho_h=1$, $\rho_H=2$, $\epsilon=0.1$, $\beta = \pi/4$.
}
\end{figure}

\begin{figure}[!htb]
\centering
\leavevmode\epsfxsize=10.0cm
\epsfbox{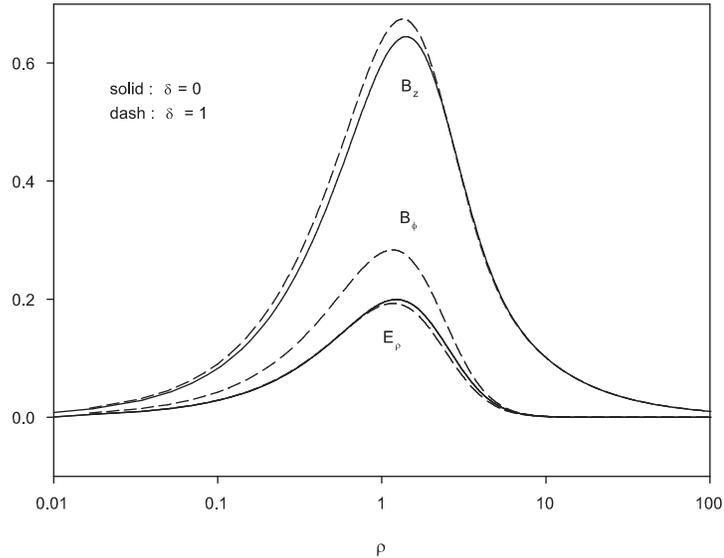}\\
\caption{\label{electrom}
 The profile of the electromagnetic fields associated with an embedded solution (solid) and 
 a generic doubly twisted solution (dashed) are compared
}
\end{figure}

 We manage to construct doubly-twisted solutions for generic values of the potential, i.e. not NO-embedded solutions.
 From now on we will use the parametrisation of the coupling constants formulated in terms of 
 $\rho_H,\rho_h$ and we will concentrate on the case $\rho_h=1, \rho_H=2$, $\beta= \pi/4$; corresponding to the
 physical region for 
 $-3/2 < \epsilon_3 < 3/2$. 
 The profile of  a doubly twisted solution is reported on Fig. \ref{profile}. 
 Let us stress that the functions $\Delta$ and $a_3$
 are {\it not} proportional to each other for generic doubly twisted solutions. 
 The non vanishing components of the electromagnetic fields $\vec E$ and $\vec B$ read
 \be
      E_{\rho} = - \frac{dA_0}{d\rho} \ \ , \ \ B_{\phi} = - \frac{dA_z}{d\rho} \ \ , \ \ 
      B_z = \frac{d A_{\phi}}{d \rho} + \frac{A_{\phi}}{\rho}
 \ee
 These functions are represented on Fig.\ref{electrom} for an embedded solution (solid lines) and for the doubly twisted solution of Fig. \ref{profile}. Note that  the 
 embedded solution has    $E_{\rho} = B_{\phi}$ since  $\Delta(\rho)=0$. 
 

\begin{figure}[!htb]
\centering
\leavevmode\epsfxsize=10.0cm
\epsfbox{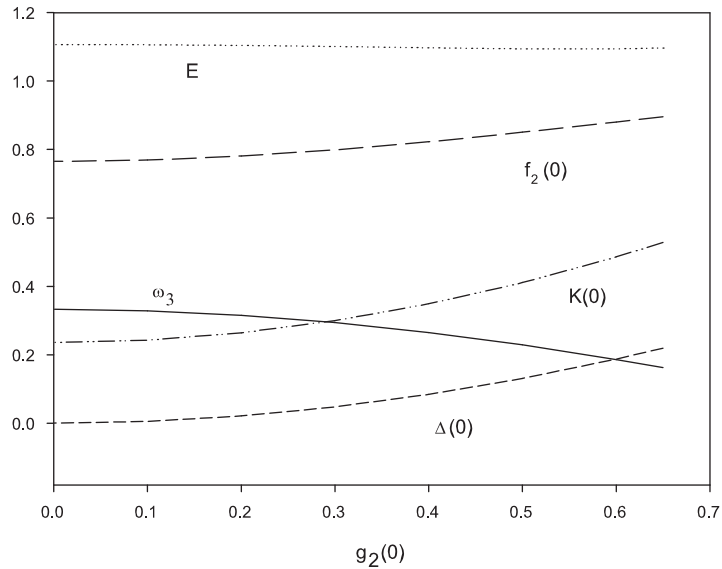}\\
\caption{
\label{fig_03} 
The evolution of some data for a doubly twisted solution with $R_0 = 2, R_3 = -0.2$ and for
$\rho_H=2, \rho_h=1$, and $\epsilon=0.1$, $\beta = \pi/4$.
}
\end{figure}

\begin{figure}[!htb]
\centering
\leavevmode\epsfxsize=10.0cm
\epsfbox{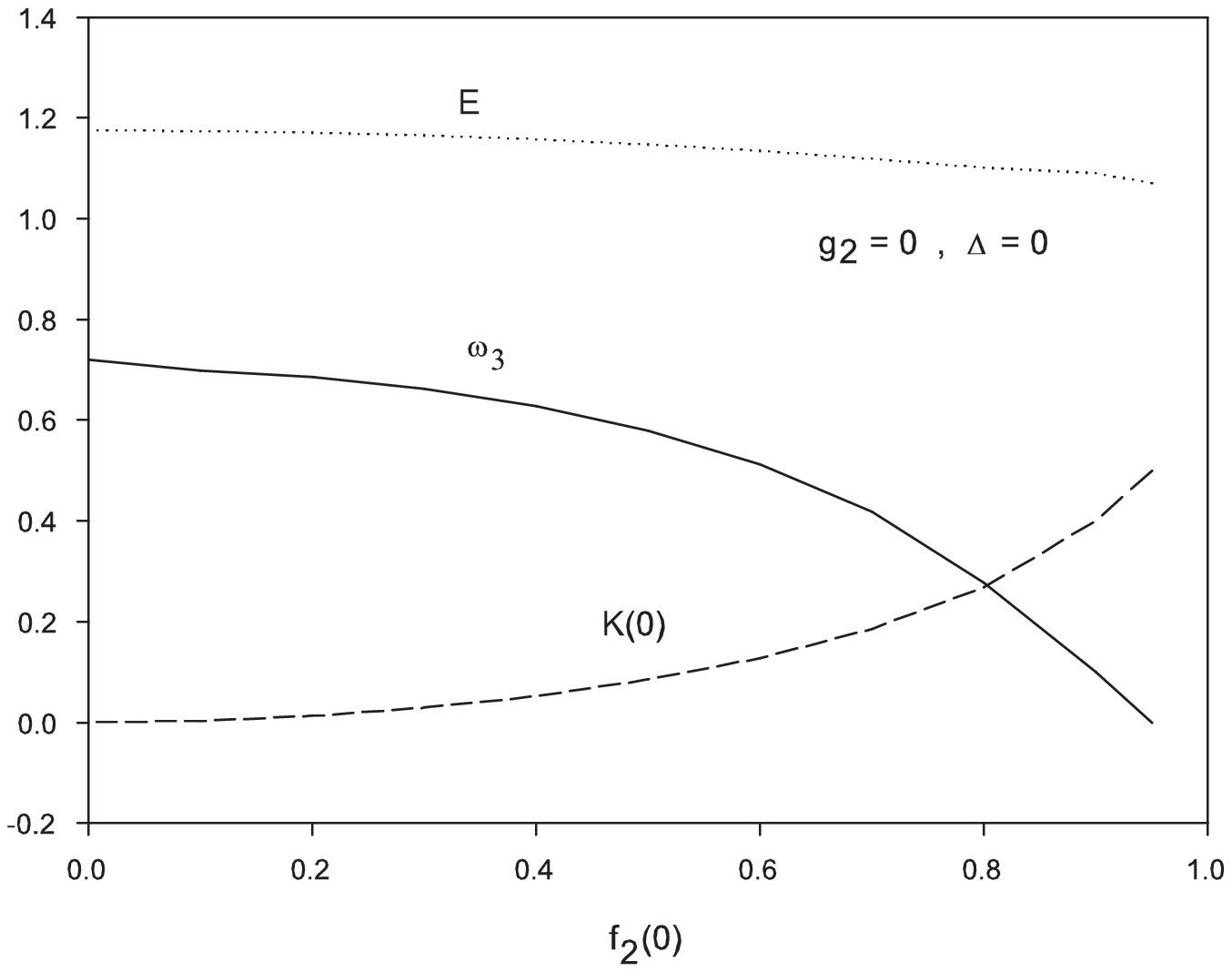}\\
\caption{
\label{fig_04} 
The evolution of some data for a simply twisted solution with $R_0 = 2, R_3 = -0.2$ and for
$\rho_H=2, \rho_h=1$, and $\epsilon_3=0.1$, $\beta = \pi/4$.
}
\end{figure}

 Several features of the stationary twisted solutions are summarized on Fig.\ref{fig_03} and Fig. \ref{fig_04}. 
 On Fig. \ref{fig_03} 
 some parameters characterising the evolution of a doubly-twisted  solution 
 ($f_2(0) \neq 0$, $g_2(0) \neq 0$) into a simply-twisted solution (in the limit $g_2(0) \to 0$) are presented.
 Although the effect on the energy is rather small, we have checked numerically that the energy
 slightly decreases while $g_2(0)$ increases. Finally, on Fig. \ref{fig_04}, the evolution of the limiting
 solution of Fig.\ref{fig_03} for $g_2(0)=0$ is reported for decreasing $f_2(0)$. As expected, the energy of the
 twisted solution is lower than the one of the untwisted one. The critical value  $\omega_b$ can be
 read on the picture. Although it would need more checks, these results suggest that doubly twisted
 string can have lower classical energy than the simply-twisted  and the untwisted ones. A systematic check
 of this statement would need more numerical and/or analytical work.

 \begin{figure}[!htb]
\centering
\leavevmode\epsfxsize=10.0cm
\epsfbox{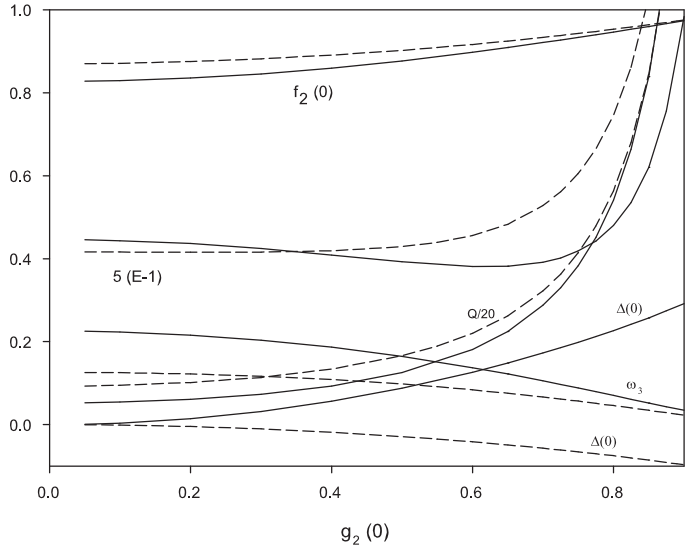}\\
\caption{
\label{noel3} 
The evolution of some parameters as functions of the value $g_2(0)$ for $R_0=1$ and $R_3=-0.5$ (solid)
and $R_3=1.5$ (dashed)
}
\end{figure} 
 
 Further aspects of the doubly twisted solutions is presented on Fig. \ref{noel3}. Here some features
 of  the solutions are superposed for two values of the parameter $R_3$ with all other parameters
  fixed (namely $R_0=1$, $\omega_0=0.01$) and $g_2(0)$ labels the branch of solutions. The solid and dashed
  curves correspond to $R_3 = -0.5$ and $R_3 = 1.5$ respectively.
  
 In the limit $g_2(0) \to 0$ the branch bifurcates into a simply twisted solution. 
 In the limit
 $g_2(0) \to 1$ it ends up into a non localized solution. This figure also confirms the property that
 doubly twisted solutions have a lower energy than the simply twisted ones at least close to the bifurcating
 point where $g_2(0)=0$. In fact the decrease of the enegy is very small for the dashed curve but more
 significant for $R_3 < 0$ corresponding to the solid curve. After reaching a minimal value, the 
 energy increases with $g_2(0)$. 

 \begin{figure}[!htb]
\centering
\leavevmode\epsfxsize=10.0cm
\epsfbox{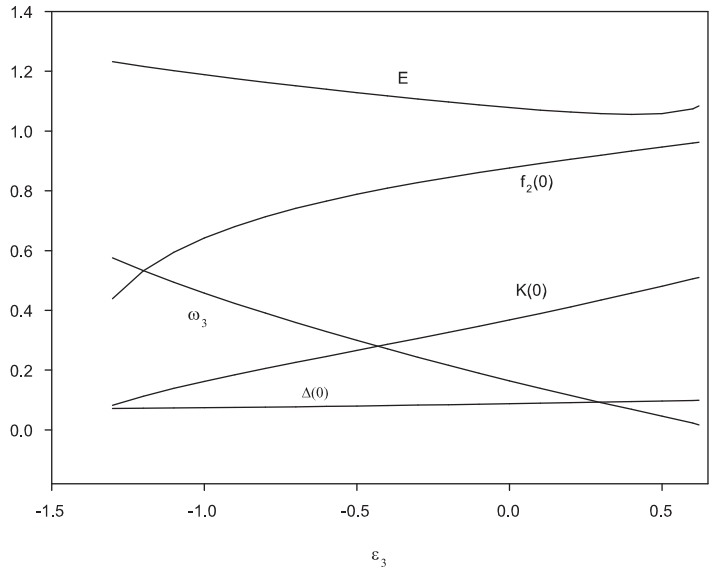}\\
\caption{
\label{noel2} 
The evolution of some parameters as  functions of the coupling constant $\epsilon_3$ for $R_0=1$ and $R_3=-0.5$
}
\end{figure} 
So far, all figures are presented with fixed value of $\epsilon_3=0$. However, as suggested in \cite{brihaye}, 
this coupling constant plays a role in the study of bifurcations. The analysis of the full bifurcation pattern
for varying $\epsilon_3$ (see Fig. 1 of  \cite{brihaye})
in the present context is out of the scope of this paper due to the richness of the
solution's manifold. Some information about the response of doubly-twisted solutions to $\epsilon_3$
is presented on Fig. \ref{noel2}. This graphic is obtained for $R_0=1, R_3=-0.5$, $g_2(0)=0.5$ and $\rho_h=1, \rho_H=2$,
$\beta = \pi/4$
(corresponding to the physical domain $-3/2 < \epsilon_3 < 3/2$).
For increasing $\epsilon_3$, finite energy solutions stop to exist for $\epsilon_3 \sim 0.62$. In this limit
we observe that $\omega^2 = 0$; as a consequence the various functions  stop to be localized. For negative values
of $\epsilon_3$, the numerical analysis becomes very involved for $\epsilon_3 \to -3/2$. Our results
strongly suggest that, in this limit, the solution ends up into a simply twisted solution with respect to the doublet
$\phi_{(2)}$, i.e. it has $f_2(\rho)=0$ and $g_2(\rho)\neq 0$. 
The fact that $\Delta(\rho) \neq 0$ in this limit is related to the 
parametrisation (\ref{ansatz}) adopted for this field in terms of $\vec \omega$.
 
\subsection{Oscillating solution}  
\begin{figure}[!htb]
\centering
\leavevmode\epsfxsize=10.0cm
\epsfbox{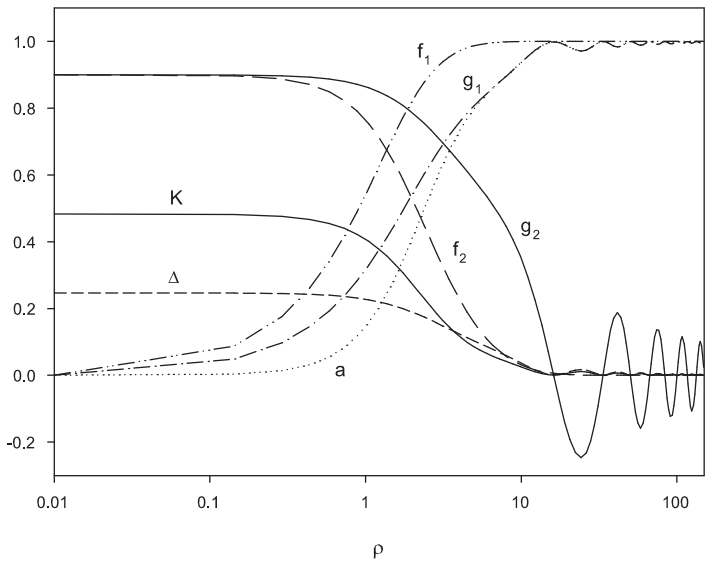}\\
\caption{
\label{oscillo} 
An example of oscillating solution with $R_0 = 1, R_3 = -0.19$, $\omega_0=0.2$, $\omega_3 = 0.3009$ and for
$\rho_H=2, \rho_h=1$, and $\epsilon=0.1$, $\beta = \pi/4$.
}
\end{figure}

In \cite{fv2} it was argued that timelike (or electric, i.e. with $\omega^2 < 0$), finite energy  
solutions do not exist but no such solutions was reported. We failed
to construct such solution in the case of one doublet but in the case of two doublets, the numerical
techniques allowed us to construct such solutions. Refering to the asymptotic expansion of \cite{fv2}
it can be expected that several functions associated with timelike solutions 
 oscillate asymptotically. We obtained indeed several such solutions.
The profile for one of them is presented on Fig.\ref{oscillo}. The parameters are such that
$\sigma^2 \equiv \sigma_0^2-\sigma_3^2 < 0$, explaining in particular the oscillations of the function $g_2$.
Because this function behaves asymptotically according to $g_2 \sim \sin(i \sigma \rho)/\sqrt{\rho}$, the 
 term $\rho (g_2')^2$ in the energy density leads to an infinite energy.
\section{Conclusions}
The lagrangian considered in this paper 
admits  numerous types of physically relevant classical solutions. The most studied ones so far are 
the sphaleron \cite{ma,km} constructed by using spherically symmetric solutions in the limit $g'=0$
and with an axial symmetry  \cite{kkb} for the physical value of the Weinberg angle.
It is believed that they are related to baryon number violations processes which likely occurred copiously 
at some stages of the evolution of the Universe \cite{ru,trodden,shap}. It has been known for some time that
new solutions: the bisphaleron \cite{bk}, bifurcates from the standard sphaleron for a sufficiently
large value of the Brout-Englert-Higgs-boson mass \cite{bk,yaffe}.

The recently discovered twisted semilocal strings \cite{fv1,fv2} and 
the superconducting electroweak strings \cite{volkov} open new ways of investigations
of classical solutions in the MSSM. These solutions have applications in cosmology.
The cylindrically symmetric ansatz used in this paper transforms the formidable system of Euler-Lagrange equations
into  a 
system of seven differential equations with boundary conditions
and depending effectively on the five parameters characterizing the potential. 
The solutions are characterized by several integers labelling the winding numbers of the different
components of the scalar fields relative to the axis of the symmetry. 
The solutions present a rich pattern of bifurcations occuring at specific values of the coupling constants
which parametrize the underlying BEH-boson masses.
Solutions with $\Delta=a_3=f_2=g_2=0$ (NO-type) likely exist for all values of the coupling constants.
However for sufficiently
high values of the BEH-boson masses, solutions  bifurcate from the NO-type solution and new solutions
exist. They depend on time through a non trivial phase
factor and can be made z-depending (denoting $z$ as the coordinate associated with the symmetry axes)
by a Lorentz boost in this direction. Very reminiscent of the
bisphaleron-sphaleron bifurcation, the static twisted strings have lower energy than the untwisted one.
In contrast to the bisphaleron,  which exist for $M_H \sim 12 M_W$,
twisted semilocal strings arise in regions of the parameter space completely compatible with the experimental lower bounds
of the BEH-boson masses.
\small{
 
}
\end{document}